%% file: main.tex
\documentclass[sigconf,natbib=true,anonymous=false]{acmart}
\AtBeginDocument{%
  \providecommand\BibTeX{{%
    \normalfont B\kern-0.5em{\scshape i\kern-0.25em b}\kern-0.8em\TeX}}}

\setcopyright{acmcopyright}
\copyrightyear{2023}
\acmYear{2023}
\acmDOI{XXXXXXX.XXXXXXX}

\acmConference[Conference acronym 'XX]{Make sure to enter the correct
  conference title from your rights confirmation emai}{June 03--05,
  2018}{Woodstock, NY}
\acmPrice{15.00}
\acmISBN{978-1-4503-XXXX-X/18/06}

\usepackage{amsmath,amssymb,amsfonts}

\usepackage{algorithmic}
\usepackage{graphicx}
\usepackage{textcomp}
\usepackage{xcolor}
\usepackage{multirow}
\usepackage{subfigure}
\usepackage{balance}
\usepackage{booktabs}
\usepackage{threeparttable}
\usepackage[ruled,linesnumbered]{algorithm2e}

\newcommand{\nosection}[1]{\vspace{2pt}\noindent\textbf{#1.}}

\newcommand{\M}{In-UCDS}
\newcommand{\Mfirst}{UCDS}
\newcommand{\UOFMatric}{$\mathcal{M}_{UOF}$}

\copyrightyear{2023}
\acmYear{2023}
\setcopyright{acmlicensed}\acmConference[MM '23]{Proceedings of the 31st
ACM International Conference on Multimedia}{October 29-November 3,
2023}{Ottawa, ON, Canada}
\acmBooktitle{Proceedings of the 31st ACM International Conference on
Multimedia (MM '23), October 29-November 3, 2023, Ottawa, ON, Canada}
\acmPrice{15.00}
\acmDOI{10.1145/3581783.3613831}
\acmISBN{979-8-4007-0108-5/23/10}

\begin{document}

\title{In-processing User Constrained
Dominant Sets for User-Oriented Fairness in Recommender Systems}

\author{Zhongxuan Han}
\affiliation{
\institution{College of Computer Science and Technology, Zhejiang University}
\country{Hangzhou, China}
}
\email{zxhan@zju.edu.cn}

\author{Chaochao Chen}
\authornote{Chaochao Chen is the corresponding author.}
\affiliation{
\institution{College of Computer Science and Technology, Zhejiang University}
\country{Hangzhou, China}
}
\email{zjuccc@zju.edu.cn}

\author{Xiaolin Zheng}
\affiliation{
\institution{College of Computer Science and Technology, Zhejiang University}
\country{Hangzhou, China}
}
\email{xlzheng@zju.edu.cn}

\author{Weiming Liu}
\affiliation{
\institution{College of Computer Science and Technology, Zhejiang University}
\country{Hangzhou, China}
}
\email{21831010@zju.edu.cn}

\author{Jun Wang}
\affiliation{
\institution{OPPO Research Institute}
\country{Shenzhen, China}
}
\email{junwang.lu@gmail.com}

\author{Wenjie Cheng}
\affiliation{
\institution{College of Computer Science and Technology, Zhejiang University}
\country{China}
}
\email{wenjiec@zju.edu.cn}

\author{Yuyuan Li}
\affiliation{
\institution{College of Computer Science and Technology, Zhejiang University}
\country{Hangzhou, China}
}
\email{11821022@zju.edu.cn}

\renewcommand{\shortauthors}{Zhongxuan Han et al.}

\input{chapters/abstract}

\keywords{Recommender System, Fairness, Dominant Sets}

\ccsdesc[500]{Information systems~Recommender systems}

\maketitle

\input{chapters/introduction}

\input{chapters/related_work}
\input{chapters/preliminary.tex}

\input{chapters/the_proposed_framework.tex}

\input{chapters/experiments_and_analysis.tex}

\input{chapters/conclustion.tex}

\begin{acks}
This work was supported in part by the National Natural Science Foundation of China (No. 72192823), the “Ten Thousand Talents Program” of Zhejiang Province for Leading Experts (No. 2021R52001).
\end{acks}

\bibliographystyle{ACM-Reference-Format}
\balance
\bibliography{reference}


\clearpage
\input{chapters/appendix.tex}

\end{document}

%% file: chapters/abstract.tex
\begin{abstract}
Recommender systems are typically biased toward a small group of users, leading to severe unfairness in recommendation performance, i.e., User-Oriented Fairness (UOF) issue.
The existing research on UOF is limited and fails to deal with the root cause of the UOF issue: the learning process between advantaged and disadvantaged users is unfair.
To tackle this issue,
we propose an \textbf{In}-processing \textbf{U}ser \textbf{C}onstrained \textbf{D}ominant 
 \textbf{S}ets (\M) framework, which is a general framework that can be applied to any backbone recommendation model to achieve user-oriented fairness.
We split \M~into two stages, i.e., the \textit{\Mfirst~modeling stage} and the \textit{in-processing training stage}.
 In the \Mfirst~modeling stage, for each disadvantaged user, we extract a constrained dominant set (a user cluster) containing some advantaged users that are similar to it.
 In the in-processing training stage, we move the representations of disadvantaged users closer to their corresponding cluster by calculating a fairness loss.
 By combining the fairness loss with the original backbone model loss, we address the UOF issue and maintain the overall recommendation performance simultaneously.
 Comprehensive experiments on three real-world datasets demonstrate that \M~outperforms the state-of-the-art methods, leading to a fairer model with better overall recommendation performance.

\end{abstract}

%% file: chapters/introduction.tex
\section{Introduction}
\begin{figure}[t]
    \centering
    \subfigure[Gradient Norm Distribution]{
		\label{introduction-norm}
		\includegraphics[width=0.45\linewidth]{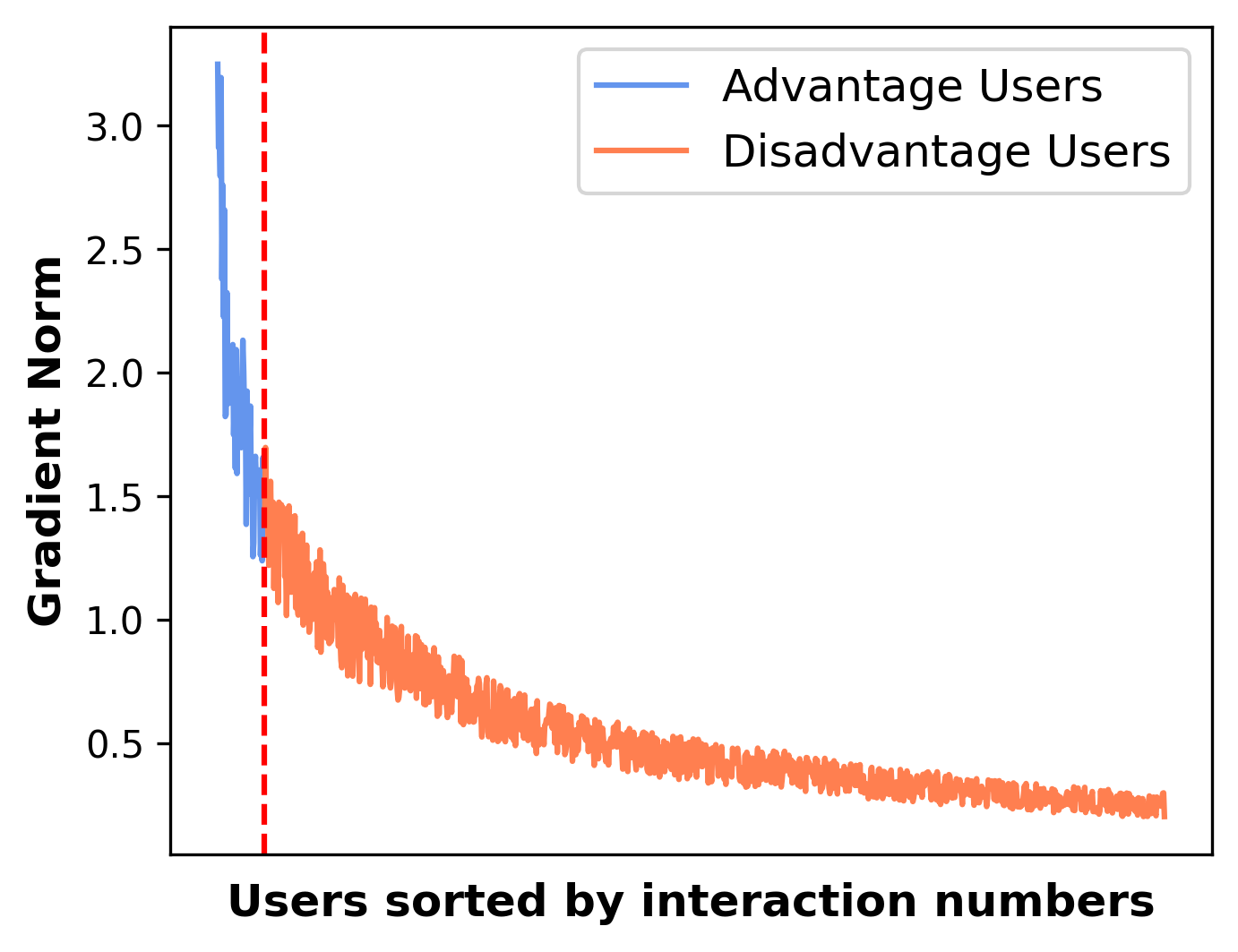}}
    \subfigure[Performance Differences]{
		\label{introduction-performance}
		\includegraphics[width=0.5\linewidth]{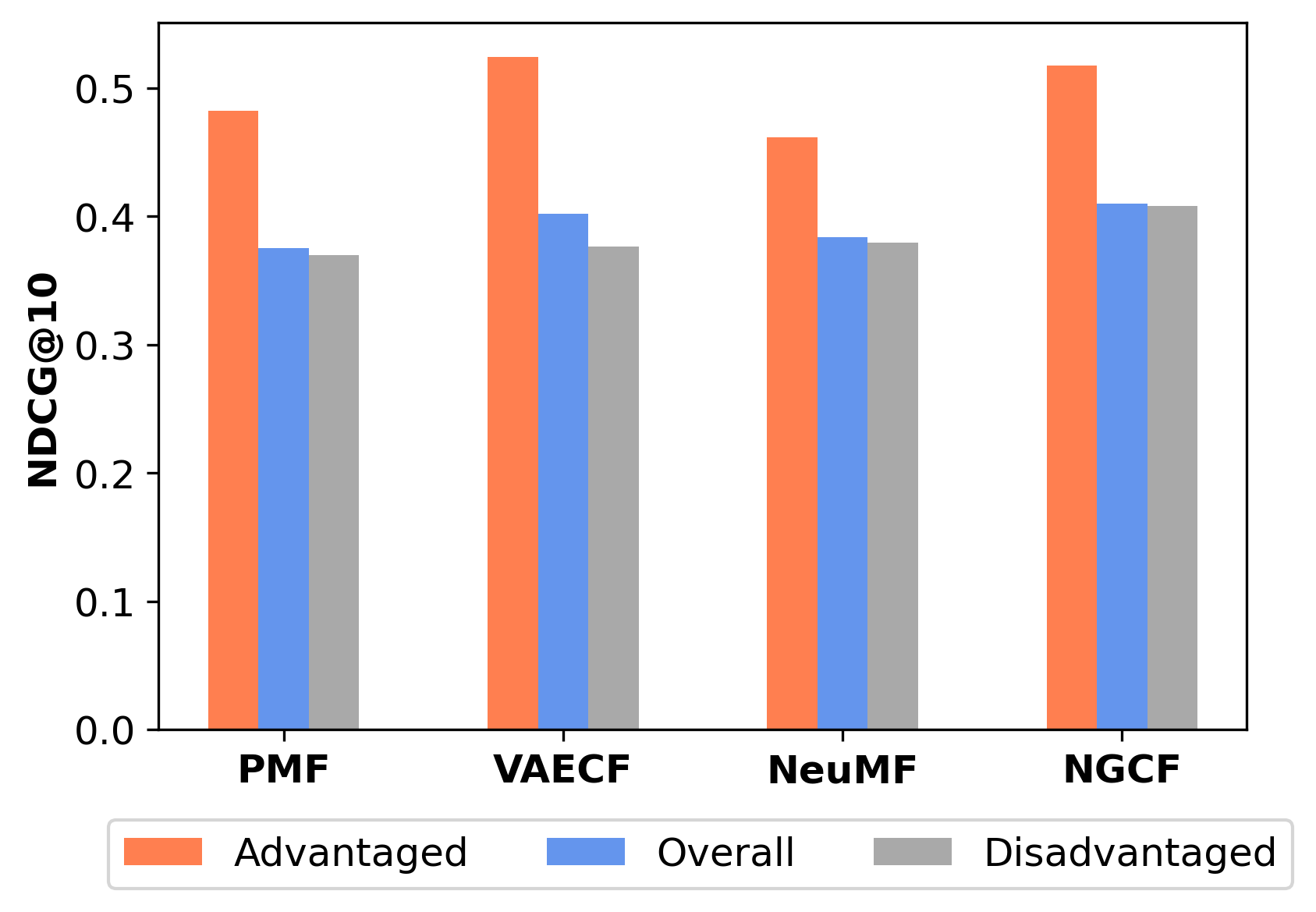}
		}
   \vspace{-10pt}
    \caption{(a) and (b) show the results obtained from experiments on Gowalla \cite{liu2017experimental} dataset. (a) visualizes the norm (i.e., $L_2-norm$ ) of gradients coming from different users in a training epoch of NeuMF \cite{he2017neural}. A bigger gradient norm indicates a bigger contribution to the updating of the model. (b) shows the significant fairness issue between advantaged users and disadvantaged users. Advantaged users can always get much better recommendation results. 
    }
    \label{introduction-figure}
    \vspace{-2pt}
\end{figure}

Nowadays, fairness has become an essential part of the machine learning community \cite{dai2022comprehensive, mehrabi2021survey, binns2018fairness, hutchinson201950, verma2018fairness}, and is also widely studied in Recommender Systems (RSs) \cite{deldjoo2022survey}.
RS is a complex field involving frequent interactions between users and items~\cite{burke2017multisided, 10.1145/3543507.3583402,10.1145/3543507.3583366,10.1145/3503161.3548072}.
Fairness issues are common on both
the user's side \cite{li2021user, rahmani2022experiments} and the items' side \cite{deldjoo2021flexible, dash2021umpire}.
%
In this paper, we consider the fairness issues of the performance divergence regarding different user groups.

\input{tables/interactions.tex}

The fairness issue about different groups of users is essential in RSs.
RSs are always biased toward a small group of users, leading to severe unfairness in recommendation performance~\cite{li2021user, rahmani2022experiments, hao2021pareto}.
We define these users with more satisfied recommendation results as \textbf{advantaged users} and other users as \textbf{disadvantaged users}.
According to~\cite{li2021user, rahmani2022experiments}, users' different activity levels (e.g., interactions, total consumptions, and maximum consumption amount) will have different impacts on the training process of the recommendation models, which leads to unfair treatments.
%
%
In this paper, we take users' interaction numbers as an example factor that leads to the fairness issue.
%
%
According to Figure~\ref{introduction-norm}, advantaged users (mostly with more interactions) always contribute more model updating when training a recommendation model.
Thus, as shown in Figure~\ref{introduction-performance}, the model is biased towards or even dominated by those advantaged users, leading to a significant fairness issue between advantaged and disadvantaged users.
What is more, the majority of users in a recommendation scenario are always those disadvantaged users. 
As shown in Table~\ref{interaction-table}, the percentage of users who interact frequently (i.e., advantaged users) is relatively small. 
Thus the above bias between advantaged and disadvantaged users will reduce overall user satisfaction.
Here, we call the unfair treatment between advantaged and disadvantaged users as \textbf{U}ser-\textbf{O}riented \textbf{F}airness (UOF)~\cite{li2021user, rahmani2022experiments} issue.

\begin{figure}[t]
  \centering
  \includegraphics[width=\linewidth]{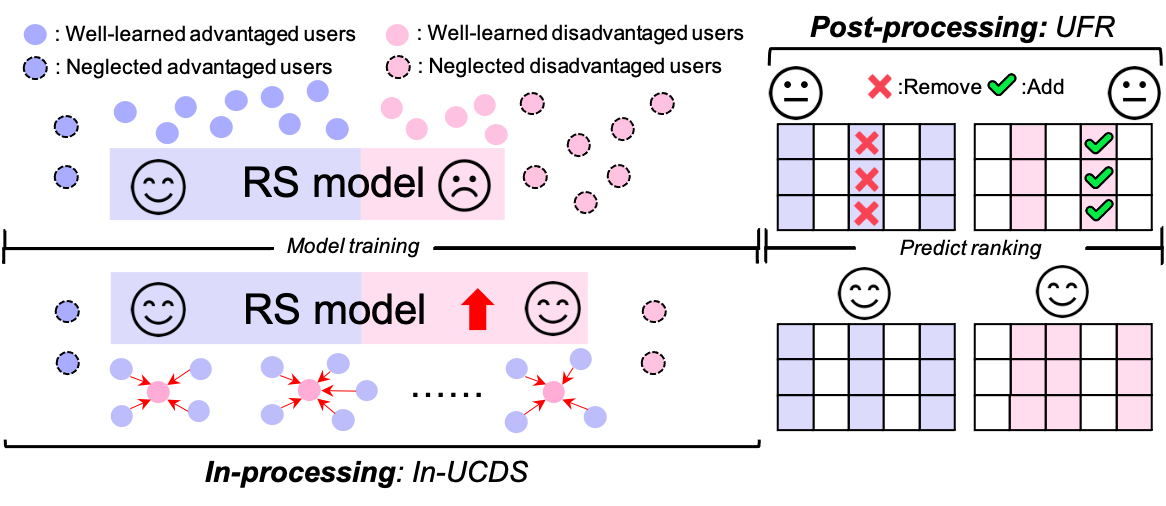}
  \vspace{-20pt}
      \caption{\M~comparing with UFR. Recommender models always neglect more disadvantaged users. \M~let disadvantaged users learn from similar advantaged users to enhance the learning process of disadvantaged users.}
      \label{comparison-figure}
      \vspace{-20pt}
\end{figure}

%
To date, the relevant work of UOF is quite limited.
Basically, existing work solving the fairness issue can be divided into \textit{pre-processing} methods (solving the bias in training data), \textit{in-processing} methods (combing the model training process with fairness consideration), and \textit{post-processing} methods (directly modifying the recommendation results to achieve fairness) \cite{dai2022comprehensive}.
%
%
Previous studies~\cite{li2021user,rahmani2022experiments} have examined the issue of UOF and presented the User-oriented Fairness Re-ranking (UFR) method, which serves as a post-processing solution for addressing UOF concerns.
As shown in Figure~\ref{comparison-figure}, UFR only modifies the recommendation result \textit{after} model training and fails to alleviate the training bias between advantaged and disadvantaged users.
%
%
Figure~\ref{introduction-norm} illustrates that advantaged users always contribute more to model training.
Thus, \textit{the root cause} of the UOF issue is the bias between advantaged and disadvantaged users during model training, which cannot be addressed by the post-processing UFR approach.
%
%
The recent S-DRO approach~\cite{10.1145/3485447.3512255} proposed an in-processing solution to mitigate loss for the disadvantaged user group. 
Nonetheless, achieving substantial improvements is challenging due to the inherent limitations of training data for this group. 
In essence, prior research cannot provide a fundamental solution for addressing the UOF issue.

In this paper, we propose an \textbf{In}-processing \textbf{U}ser \textbf{C}onstrained \textbf{D}ominant \textbf{S}ets (\M) framework to give disadvantaged users a fairer learning result, to better solve the UOF issue at the fundamental level.
%
%
As shown in Figure~\ref{comparison-figure}, different from UFR, by giving fair consideration during the model training, we can solve the UOF issue  in a deeper and more natural way, i.e., correcting the bias between advantaged and disadvantaged users in the training process.
The fundamental concept behind \M~is to identify a specific number of advantaged users who share similarities with each disadvantaged user. 
This approach enables disadvantaged users to learn from the well-trained advantaged users during model training, rather than relying solely on their limited training data, as UFR and S-DRO do.
By doing so, the learning gaps between these two user groups can be significantly reduced.
To achieve this goal, there are mainly two key problems: (1) how to find similar advantaged users for each disadvantaged user; (2) how to design a method to let each disadvantaged user learn from similar advantaged users.

\M~consists of two stages, i.e., the \textit{\Mfirst~modeling stage} and the \textit{in-processing training stage}.
Each stage is respectively devised for one of the aforementioned problems.
%
(1) In the \Mfirst~modeling stage, we design a user constrained dominant sets approach to find a set of similar advantaged users for each disadvantaged user.
Constrained dominant sets~\cite{zemene2016interactive, alemu2019deep} can explore a cluster of the most similar nodes for a particular node in a graph, which is naturally suitable for our task.
Thus, we treat the \Mfirst~modeling stage as finding an advantaged user cluster with a given disadvantaged user as a constraint.
Specifically, for a given user-item interaction graph, \textit{firstly}, we build a user-user graph for each disadvantaged user, which includes all advantaged users.
To give an initial relationship among these user nodes, we set the weight of each edge between two user nodes as the number of items they both interact with.
Thus a bigger weight indicates a closer connection.
\textit{Secondly}, we treat each disadvantaged user as a target user to find a constrained cluster in each graph so that we can find the most similar advantaged user set for each disadvantaged user.
(2) In the in-processing training stage, we define a fairness loss that calculates the representation difference between each disadvantaged user and users in its constrained cluster.
By minimizing the fairness loss, the representations of each disadvantaged user will move closer to those of similar advantaged users, leading to improved learning results during model training.
Through the above two stages, \M~will narrow the representation gap between advantaged and disadvantaged users.
By incorporating the fairness loss with the loss of the original backbone model, \M~enhance fairness without sacrificing the recommendation performance.

To evaluate the performance of our proposed \M~framework, we conduct extensive experiments on three real-world datasets based on four different backbone recommendation models.
Multiple evaluation metrics demonstrate that \M~outperforms the State-Of-The-Art (SOTA) models UFR and S-DRO in all backbone recommendation models from various perspectives.
\M~can not only mitigate fairness issues among users of different activity levels, but also bring a much higher overall performance, proving that we can handle the UOF issue more effectively.

We summarize our main contributions as follows: (1) We first propose an in-processing framework \M~to solve the fairness issues between advantaged and disadvantaged users in RS.
By doing so, we can alleviate the bias of RS among users with different activity levels and bring better recommendation performance.
(2) We design a constrained dominant sets approach to find an advantaged user cluster for each disadvantaged user to enhance its representation during training.
(3) We conduct extensive experiments on three real-world datasets, demonstrating that \M~outperforms the SOTA methods from various perspectives.
%

%% file: tables/interactions.tex
\begin{table}[t]
\renewcommand{\arraystretch}{1}
\vspace{-5pt}
   \caption{Percentage of users with different numbers of interactions ($n$).}
  \label{interaction-table}
  \vspace{-10pt}
  \begin{threeparttable}
  
    \begin{tabular}{ccccc}
    \hline
    Dataset & $n \ge 30$ & $n \ge 50$ & $n \ge 80$ & $n \ge 100$ \\ \hline
    
    Epinion & 50.39\% & 18.38\% & 6.35\% & 3.14\%\\ \hline
    
    MovieLens & 86.91\% & 69.62\% & 53.24\% & 46.06\% \\ \hline

    Gowalla & 36.03\% & 15.14\% & 4.74\% & 2.30\% \\ \hline
    \end{tabular}
    \end{threeparttable}
    \vspace{-15pt}
\end{table}

%% file: chapters/related_work.tex
\vspace{-5pt}
\section{Related Work}
%

\input{chapters/related_work/Algorithmic_Fairness}

\input{chapters/related_work/Fair_Recommendation.tex}

\input{chapters/related_work/Dominant_Set.tex}

%% file: chapters/related_work/Algorithmic_Fairness.tex
\subsection{Fairness in Machine Learning}
Fairness has become one of the most important topics in machine learning.
According to \cite{mehrabi2021survey}, fairness in a decision-making scenario can be defined as the absence of any prejudice or favoritism toward an individual or a group based on their inherent or acquired characteristics.
From different perspectives, research on fairness in machine learning can be divided into different categories.

\textit{Regarding the groups affected by fairness issues}, the research about fairness can be divided into two aspects, i.e., group fairness and individual fairness \cite{mehrabi2021survey, dai2022comprehensive}.
Group fairness aims to treat users of different groups equally.
Related work includes Equalized Odds \cite{hardt2016equality}, Equal Opportunity \cite{hardt2016equality}, Conditional Statistical Parity \cite{corbett2017algorithmic}, Demographic Parity \cite{kusner2017counterfactual}, and Treatment Equality \cite{berk2021fairness}.
Individual fairness aims to give similar individuals similar recommendation results.
Related work includes Fairness Through Unawareness \cite{grgic2016case}, Fairness Through Awareness \cite{dwork2012fairness}, and Counterfactual Fairness~\cite{kusner2017counterfactual}.

\textit{Regarding the fairness algorithms acting on the different stages of the backbone model}, the research about fairness can be divided into three aspects, i.e., pre-processing methods, in-processing methods, and post-processing methods \cite{mehrabi2021survey, dai2022comprehensive}.
Pre-processing methods try to transform the training data so that the underlying discrimination can be removed before training the model~\cite{d2017conscientious, kang2020inform}.
In-processing methods aim to add fairness consideration into the training stage of the state-of-the-art model to remove discrimination during the training~\cite{dai2020learning, bose2019compositional}.
Post-processing methods directly modify the recommendation results of a given model after training \cite{li2021user, kang2020inform}.

In this paper, we propose an in-processing framework to solve the rarely studied user-oriented fairness problem, a kind of group fairness issue, in recommender systems.

%% file: chapters/related_work/Fair_Recommendation.tex
\subsection{Fair Recommendation}
A recommender system is a complex field with multiple stakeholders~\cite{burke2017multisided,li2023selective,10.1145/3572407,chen2022differential}.
The fairness issue is prevalent on providers' side, items' side, and users' side, introducing a trade-off among different roles.

\textit{For fairness on providers' side.} 
The length of time cooperating with different providers~\cite{gharahighehi2021fair, ferraro2019music}, the correspondence between providers and items~\cite{boratto2021interplay}, and some characteristics of providers~\cite{shakespeare2020exploring} are all factors that contribute to fairness issues.
%
%
\textit{For fairness on items' side.}
Sensitive attributes (e.g., price, brand, and geographical area)~\cite{deldjoo2021flexible, burke2018balanced} and popularity~\cite{wundervald2021cluster, sun2019debiasing}are key factors that lead to fairness issues.
%
%
\textit{For fairness on users' side.}
The sensitive attributes of users always affect the recommendation result, e.g., gender \cite{deldjoo2021explaining, melchiorre2021investigating}, age \cite{deldjoo2021explaining}, and race \cite{gorantla2021problem}.
%

%
In this paper, we focus on the rarely studied fairness issue among users with different activity levels, i.e., the UOF problem.
Different from existing work \cite{li2021user, rahmani2022experiments, 10.1145/3485447.3512255}, we propose an in-processing method that enhances the training process of disadvantaged users to narrow the learning gap between advantaged and disadvantaged user groups.

%% file: chapters/related_work/Dominant_Set.tex
\subsection{Dominant Sets}
Dominant Sets (DS) \cite{pavan2006dominant} is a kind of graph clustering algorithm.
Zemene et al. \cite{zemene2016interactive} proposed a Constrained Dominant Sets (CDS) approach to find a cluster restrict to contain several given nodes in a graph. 
Recently, DS and CDS are widely used in computer vision applications, proving their effectiveness in different kinds of clustering jobs.
Zemene et al. \cite{zemene2018dominant} presented CDS with the image segmentation task.
Wang et al. \cite{wang2019dominant} combine the dominant sets clustering process in a recursive manner to select representative images from a collection of images for the 3D object recognition task.
Alemu et al. \cite{alemu2019deep} proposed an end-to-end constrained clustering scheme to tackle the person re-identiﬁcation (reid) problem. 
Different from the above methods which are widely utilized in computer vision tasks, to the best of our knowledge, in this paper we are the first to introduce CDS to the user clustering task in recommender systems.

%% file: chapters/preliminary.tex
\section{Preliminary}
In this part, we first give the problem formulation and then generally introduce dominant sets and constrained dominant sets as the technical preparation for \M.
\subsection{Problem Formulation}
We use $\mathcal{U}$ and $\mathcal{I}$ to represent the user set and the item set.
We divide users into disadvantaged user group $\mathcal{T} = \{T_1, T_2, \dots, T_{M-1}, T_M\}$ and advantaged user group $\mathcal{A} = \{A_1, A_2, \dots, A_{N-1}, A_N\}$ based on their activity levels, where M and N indicate the number of users in each group.
According to \cite{li2021user, rahmani2022experiments}, various behaviors (i.e., interactions, total consumption, and max trade price) can all reflect users' activity levels.
Since recommendation models are always trained directly based on user-item interactions, without loss of generality, we choose users' interaction numbers (users with more interactions are more likely to be advantaged) to group users.
In this paper, we aim to narrow the gap in the recommendation performance between $\mathcal{T}$ and $\mathcal{A}$ to achieve user-oriented fairness and maintain the overall recommendation performance simultaneously.
%

User-oriented fairness is a kind of group fairness \cite{hardt2016equality, dwork2012fairness}.
The principle of group fairness is to ensure groups of users with different protected sensitive attributes should be comparably treated.
According to this idea, the definition of user-oriented fairness is defined as follows \cite{li2021user, rahmani2022experiments}:
\begin{definition}[User-Oriented Fairness (UOF)]
\begin{equation}
\label{user-oriented-fairness}
    \mathbb{E}[\mathcal{M(\mathcal{A})}] = \mathbb{E}[\mathcal{M(\mathcal{T})}].
\end{equation}
\end{definition}
In the definition of UOF, $\mathcal{M}$ indicates a metric (e.g., NDCG and F1-score) that can evaluate the recommendation performance. We use $\mathcal{M}(u)$ to represent the recommendation performance of user $u$.

UOF aims to offer users with different activity levels the same recommendation performance, which is usually impossible in real-world RS.
Thus researchers \cite{li2021user, rahmani2022experiments} always calculate the difference in average recommendation performance for different user groups to evaluate the fairness of a model:

\begin{definition}[The UOF matric (\UOFMatric)]
\begin{equation}
\label{user-oriented-fairness-cal}
    \mathcal{M}_{UOF}(\mathcal{T}, \mathcal{A}) = \bigg | \frac{1}{\lvert \mathcal{A} \rvert} \sum_{A_i \in \mathcal{A}} \mathcal{M}(A_i) - \frac{1}{\lvert \mathcal{T} \rvert} \sum_{T_i \in \mathcal{T}} \mathcal{M}(T_i) \bigg |.
\end{equation}
\end{definition}

In this paper, we use the value of \UOFMatric~to evaluate the fairness of a recommendation model. 
A smaller \UOFMatric~means a fairer algorithm that tends to treat different activity groups of users equally.

\subsection{Technical Preparation}
Dominant sets and constrained dominant sets are graph clustering methods.
Without loss of generality, we suppose a graph can be represented as $G=(\mathcal{V}, \mathcal{E}, w)$, where $\mathcal{V} = \{v1, v2, \dots, v_z\}$ indicates the vertex set with $z$ represents the number of vertices.
$\mathcal{E} \in \mathcal{V} \times \mathcal{V}$ indicates the edge set.
$w:E \xrightarrow{} R_+^*$ is the positive weight function.
The weight of an edge reflects the similarities between the linked vertices.
We represent the graph $G$ with an weighted adjacency matrix $\mathbf{H} \in R^{| \mathcal{V} | \times | \mathcal{V} |}$, which is non-negative and symmetric.
$\mathbf{H}_{ij}$ is defined as $w(i, j)$ if $(i, j) \in \mathcal{E}$, and $0$ otherwise.
Note that there are no self-loops in $G$, thus all entries on the main diagonal of $\mathbf{H}$ are zero.

\nosection{Dominant Sets}
Dominant sets is a graph clustering approach that generalizes the concept of a maximal clique to construct clusters in an edge-weighted graph.
As proved in \cite{pavan2006dominant}, we can extract a node cluster by solving the following linear-constrained quadratic optimization problem:
\begin{equation}
\label{dominant-sets-equ}
    \begin{split}  
    &\text{maximize} \ f(x) = x^{'} \mathbf{H}x,
    \\ 
    &\text{subject to} \ x \in \bigtriangleup, 
    \end{split}
\end{equation}
where $x^{'}$ indicates the transpose of vector $x$.
$\bigtriangleup$ is the standard simplex of $R^z$:
\begin{equation}
\label{triangle-equ}
    \bigtriangleup = \{x \in R^z: \sum_{i = 1}^{z} x_i = 1, \text{and} \ x_i \ge 0 \ \text{for all} \ i = 1, 2, \dots, z\}.
\end{equation}

According to $x$, we can extract how relevant a node is to the cluster.
An effective way to solve the optimization above is the so-called \textit{replicator dynamics} \cite{pavan2006dominant}, which is defined as follows:
\begin{equation}
    \label{replicator-dynamic-equ}
    x_i^{t+1} = x_i^{t} \frac{(\mathbf{H}x^t)_i}{(x^t)^{'}\mathbf{H}x^t}.
\end{equation}

\nosection{Constrained Dominant Sets}
Constrained dominant sets is an extension of dominant sets.
The key idea is that by choosing one or more nodes as constraints, we can extract a cluster containing such user-selected elements.
The constrained dominant sets can also be defined as an optimization problem as follows:
\begin{equation}
    \label{constrained-dominant-set-equ}
    \begin{split}
        &\text{maximize} \ f_\mathcal{S}^\alpha(x) = x^{'} (\mathbf{H} - \alpha \hat{\mathbf{I
}}_\mathcal{S})x,
        \\ 
        &\text{subject to} \ x \in \bigtriangleup, 
    \end{split}
\end{equation}
where $\mathcal{S}$ indicates the user-selected elements set.
$\hat{\mathbf{I
}}_\mathcal{S}$ is the $z \times z$ diagonal matrix whose element is set to $1$ in correspondence to the vertices contained in $\mathcal{V} \backslash \mathcal{S}$ and $0$ otherwise.
Note that if $\alpha = 0$, the above optimization problem will degenerate to dominant sets.
Similar to Equation \eqref{dominant-sets-equ}, the optimization problem \eqref{constrained-dominant-set-equ} can be solved as follows:
\begin{equation}
    \label{constrainted-replicator-dynamic-equ}
    x_i^{t+1} = x_i^{t} \frac{((\mathbf{H} - \alpha \hat{\mathbf{I
}}_\mathcal{S})x^t)_i}{(x^t)^{'}(\mathbf{H} - \alpha \hat{\mathbf{I
}}_\mathcal{S})x^t}.
\end{equation}

%% file: chapters/the_proposed_framework.tex
\section{The Proposed Framework}
In this part, we will first give an overview of the proposed \M~framework, then describe the detail of the \Mfirst~modeling stage and the in-processing training stage.

\subsection{Overview}
\begin{figure*}[htbp]
\vspace{-10pt}
  \centering
  \includegraphics[width=\linewidth]{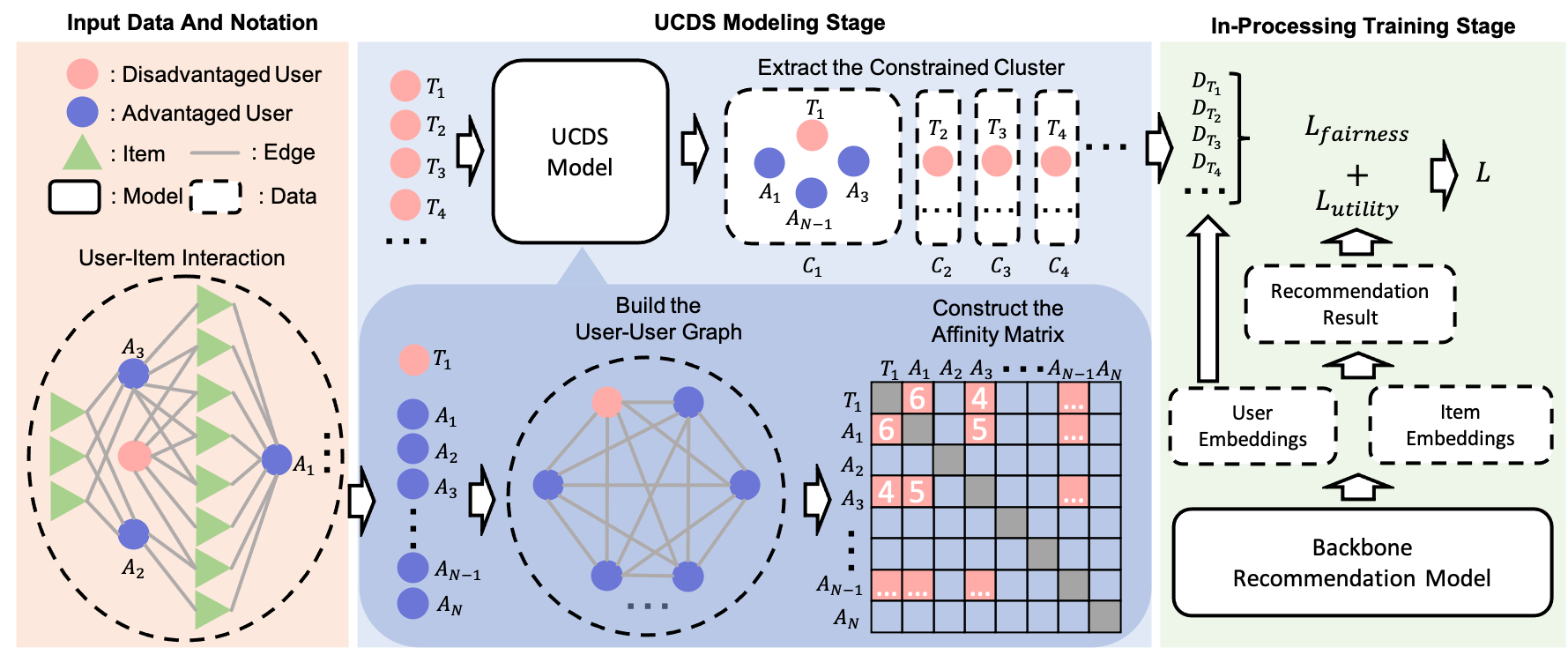}
  \vspace{-20pt}
      \caption{The overall framework. The input data and notations in this figure are shown in orange. Note that the user-item interaction graph shown in the graph is a toy example. There are many more disadvantaged users than advantaged users in real-world recommender systems. The detailed modeling stage of \Mfirst~is shown in blue, where we take the disadvantaged user $T_1$ as an example to show how to find a constrained dominant set for it in deep blue. In-processing training stage is shown in green. In each training epoch, $L_{utility}$ and $L_{fairness}$ are combined to calculate the final loss.
      }
      \label{overall-framework}
      \vspace{-10pt}
\end{figure*}

In this paper, we propose a novel \textbf{In}-processing \textbf{U}ser \textbf{C}onstrained \textbf{D}ominant \textbf{S}ets framework, namely \M, to better solve the user-oriented fairness issue in recommendation tasks.
\M~is a general framework that can combine with any recommendation models (i.e., backbone models) to achieve user-oriented fairness.
As shown in Figure \ref{overall-framework}, the overall framework is divided into two stages, i.e., the \textit{\Mfirst~modeling stage} (shown in blue) and the \textit{in-processing training stage} (shown in green).
(1) \Mfirst~can be seen as a complementary model for the backbone recommendation model.
Before the training process begins, we first extract a constrained cluster for each disadvantaged user preparing for the following in-processing training stage.
Note that the advantaged users in each cluster are similar to the corresponding target disadvantaged user identified by constrained dominant sets.
(2) During the in-processing training stage, we calculate the embedding difference between each disadvantaged user and the users in its cluster and generate the fairness loss.
We combine the fairness loss with the backbone recommendation model loss to let disadvantaged users learn from their similar advantaged users.
Through the above two stages, \M~can solve the UOF issue during the model training.

In the following sections, we will introduce the detail of the \Mfirst~modeling stage and the in-processing training stage.

\input{chapters/the_proposed_framework/user_cluster_as_constraint_dominant_set.tex}

\input{chapters/the_proposed_framework/the_training_process}

%% file: chapters/the_proposed_framework/user_cluster_as_constraint_dominant_set.tex
\subsection{\Mfirst~Modeling Stage}
The main idea of the \Mfirst~modeling stage is to find a similar advantaged user cluster for each disadvantaged user, papering for the following in-processing training stage.
Firstly, we will build a user-user graph containing all advantaged users for each disadvantaged user to give a one-to-one similarity relationship between each user pair.
Then, we will construct an affinity matrix for the graph to give the target disadvantaged user a constraint so that we can extract a cluster containing the target user.
Finally, we extract a constrained dominant set for each disadvantaged user based on the affinity matrix built above.

\nosection{Build the User-User Graph}
To explore advantaged user clusters for each disadvantaged user, we need to build a user-user graph for each target disadvantaged user first.
According to the blue part of Figure \ref{overall-framework}, each user-user graph contains one of the disadvantaged users and all advantaged users.
The graph is a complete graph with the weights of edges representing the similarities between the linked nodes.
To better find the correlations among users, we set the weight between two user nodes as the number of the same items they have interacted with.
A higher weight means a higher similarity since users who interact with the same items are likely to be similar.

\nosection{Construct the Affinity Matrix}
Given the advantaged user set $\mathcal{A}$ and the disadvantaged user set $\mathcal{T}$, each time we will build a complete user-user graph containing one of the users belonging to $\mathcal{T}$ (take $T_1$ as an example) and all users in $\mathcal{A}$.
We use $\mathbf{H}$ to represent this graph's adjacency matrix.
To find a constrained cluster for $T_1$, we build $\mathcal{U} = \{T_1\}$ as a constraint user set and define the affinity matrix $\mathbf{B}$ as follows:
\begin{equation}
    \label{affinity-matirx}
    \mathbf{B} = (\mathbf{H} - \alpha \hat{\mathbf{I
}}_\mathcal{U}).
\end{equation}
Note that the adjacency matrices for graphs built for each disadvantaged user are always different. 
Thus, the value of $\alpha$ is not a constant.
According to \cite{zemene2016interactive}, by setting $\alpha$ as:
\begin{equation}
    \label{alpha-value-equ}
    \alpha > \lambda_{max}(\mathbf{H}_{\mathcal{V} \backslash \mathcal
    U}),
\end{equation}
we can get a solution that necessarily contains elements of $\mathcal{U}$.
$\lambda_{max}$ indicates the largest eigenvalue of a matrix.

\nosection{Extract the Constrained Cluster}
After constructing the affinity matrix, we can extract a constrained cluster for each disadvantaged user.
To begin with, we will introduce three steps to extract a constrained cluster.
The three steps give a clustering method from measuring the correlation of two users to measuring the internal coherence of a user set.
Then, we will give the corresponding optimization problem for easy implementation.

\textit{Firstly}, to extract a constrained dominant set, we need to measure the similarity between two nodes in a graph.
Given a user graph $G=(\mathcal{V}, \mathcal{E}, w)$, we extract $\mathcal{S} \in \mathcal{V}$ as a non-empty user set and define:
\begin{equation}
    \label{phi-equ}
    \phi(\mathcal{S})(v_i, v_j) = w(i, j) - \frac{1}{| \mathcal{S} |} \sum_{k \in \mathcal{S}} w(i, k).
\end{equation}
Thus $\phi(\mathcal{S})$ measures the relative similarity between $v_i$ and $v_j$, with respect to the average similarity between $v_i$ and its neighbors in $\mathcal{S}$.

\textit{Secondly}, to explore the importance of $v_i$ to $\mathcal{S}$, we define a weight in a recursively way as follows:
\begin{equation}
    \label{dominant-set-W-equ}
    w_\mathcal{S}(i) = \begin{cases}
    1, \ & \text{if} \ |\mathcal{S}| = 1, \\
    \sum_{j \in \mathcal{S} \backslash \{i\}} \phi_{\mathcal{S} \backslash \{ i \}}(j, i) w_{\mathcal{S} \backslash \{ i \}}(j), \  & \text{otherwise.}
    \end{cases}
\end{equation}
A positive $w_\mathcal{S}(i)$ represents that adding $v_i$ into its neighbors in $\mathcal{S}$ will increase the internal coherence of the set, while a negative one will cause the overall coherence to be decreased.

\textit{Thirdly}, to measure the overall internal coherence of $\mathcal{S}$, we define a total weight as follows:
\begin{equation}
    \label{dominant-set-total-weight-equ}
    W(\mathcal{S}) = \sum_{i \in \mathcal{S}} w_\mathcal{S}(i).
\end{equation}
If the obtained $\mathcal{S}$ is a dominant set (i.e., a good user cluster), it should satisfy two conditions.
\begin{itemize}
\item \textbf{Condition 1}. $w_\mathcal{S}(i) > 0,$ for all $i \in \mathcal{S}$, representing a dominant set is internally coherent.

\item \textbf{Condition 2}. $w_{\mathcal{S} \cup \{i\}}(i) < 0,$ for all $i \notin \mathcal{S}$, representing that the internal coherence will be destroyed by the addition of any vertex from outside.
\end{itemize}

The above three steps give us a method to extract a constrained dominant set $\mathcal{S}$ from graph $\mathcal{G}$.
Initially we set $\mathcal{S} = \mathcal{V}$ and then drop all vertices $v_i$ with $w_{\mathcal{S}}(i) < 0$ and save all vertices $v_j$ with $w_\mathcal{S}(j) > 0$.
We give a toy example in Appendix~\ref{appendix-example} to describe how we extract a dominant set in detail.

%
Now, we give a corresponding optimization problem to effectively implement the above steps and extract a constrained dominant set.
As proved in \cite{pavan2006dominant}, for a target disadvantaged user $T_1$, the optimization problem is defined as:
\begin{equation}
    \label{user-cluster-equ}
    \begin{split}
        &\text{maximize} \ f_\mathcal{S}^\alpha(x) = x^{'}\mathbf{B}x,
        \\ 
        &\text{subject to} \ x \in \bigtriangleup.
    \end{split}
\end{equation}
To solve the optimization problem above, we just need to solve the replicator dynamics problem following Equation \eqref{constrainted-replicator-dynamic-equ}.
Then we can extract the final constraint dominant set containing $T_1$ according to the final value of $x$.

During the \Mfirst~modeling stage, we will generate dominant sets $\mathcal{C}_1, \mathcal{C}_2, \dots, \mathcal{C}_M$ for users $T_1, T_2, \dots, T_M$, preparing for the following in-processing training stage.

%% file: chapters/the_proposed_framework/the_training_process.tex
\subsection{In-Processing Training Stage}
The basic idea of the in-processing training stage is to let disadvantaged users learn from advantaged users in their corresponding clusters.
Thus we can narrow the gap in the quality of learning results between advantaged and disadvantaged users.
To achieve this goal, as shown in the green part of Figure \ref{overall-framework}, in each training epoch, we calculate the fairness loss $L_{fairness}$ that reflects the difference between each disadvantaged user and its advantaged user cluster.
Then we combine $L_{fairness}$ with the backbone model loss $L_{utility}$.
Thus we can solve the UOF problem and maintain the recommendation performance simultaneously.

\nosection{Calculate the Fairness Loss}
During each training epoch, we use the user embeddings $\mathbf{E}$ of the backbone model to calculate the fairness loss.
For a disadvantaged user $T_i \in \mathcal{T}$, we define the representation difference between it and its constrained cluster as:
\begin{equation}
    \label{D-equ}
    D_{T_i} = \parallel \mathbf{E}_{T_i} - mean({\mathbf{E}_{A_j} | A_j \in \mathcal{C}_i}) \parallel^2,
\end{equation}
where $mean$ indicates an operation to calculate the element-wise average of a set of vectors.
However, the number of advantaged users in each constrained cluster is always different. 
To give fair treatment to each disadvantaged user and find the effect of the number of users in the user cluster, we further define a hyperparameter $k$ to select the $k$ most important users from each cluster (i.e., $\mathcal{C}_i^k$).
Note that the importance of each advantaged user is reflected by the value of the corresponding element in $x$ in the result of Equation~\eqref{constrainted-replicator-dynamic-equ}.
Then we define the final $D_i$ as:
\begin{equation}
    \label{D-equ-new}
    D_{T_i} = \parallel \mathbf{E}_{T_i} - mean({\mathbf{E}_{A_j} | A_j \in \mathcal{C}_i^k}) \parallel^2.
\end{equation}

Finally, we calculate $L_{fairness}$ as the average of $D_{T_i}$ for all disadvantaged users to improve the quality of learning results for all disadvantaged users:
\begin{equation}
    \label{D-overall-equ}
    L_{fairness} = \frac{1}{|\mathcal{T}|}\sum_{T_i \in \mathcal{T}}D_{T_i}.
\end{equation}

\nosection{Putting Together}
In each training epoch, we combine the fairness loss $L_{fairness}$ with the backbone model loss $L_{utility}$ to simultaneously solve the UOF problem and maintain the overall recommendation performance:
\begin{equation}
    \label{loss-overall}
    L = L_{utility} + \beta L_{fairness},
\end{equation}
where $L_{utility}$ represents the loss function of the backbone recommendation model, constraining the model to achieve a higher recommendation accuracy,
and $\beta$ is a hyperparameter that controls the importance of $L_{fairness}$.

%% file: chapters/experiments_and_analysis.tex
\section{Experiments And Analysis}

\input{tables/dataset_statistic.tex}

\input{tables/experiment_result.tex}

To fully evaluate the proposed \M~model.
We conduct extensive experiments on three real-world datasets to answer the following questions: \textbf{Q1}: How does \M~help the backbone recommendation model deal with the user-oriented fairness issue and enhance the recommendation performance?
\textbf{Q2}: Can \M~better deal with the fairness issue and improve the recommendation performance than existing methods?
\textbf{Q3}: Can \M~narrow the gap in the quality of the learned representations between advantaged and disadvantaged users?
\textbf{Q4}: How do important hyperparameters affect \M?
\textbf{Q5}: How the UCDS approach contributes to the performance improvement of our proposed framework?
%

%


\input{chapters/experiments_and_analysis/dataset_and_experimental_settings.tex}

\input{chapters/experiments_and_analysis/overall_comparison}

\input{chapters/experiments_and_analysis/the_change_in_l_fairness.tex}

\input{chapters/experiments_and_analysis/effect_of_hyperparameter_k}

\input{chapters/experiments_and_analysis/ablation_study}

%% file: tables/dataset_statistic.tex
\begin{table}\centering \footnotesize
\renewcommand{\arraystretch}{1.1}
  \vspace{-10pt}
  \caption{The statistics of datasets}
  \vspace{-10pt}
  \label{dataset-statistic}
  \resizebox{\linewidth}{!}{
    \begin{tabular}{cccccc}
    \hline
    Dataset & Users & Items & Interactions & Sparsity & Domain \\ \hline 
    Epinion & 2,677 & 2,060 & 103,567 & 98.12\% & Opinion \\ \hline
    MovieLens & 5,738 & 3,627 & 760,814 & 96.34\% & Movie \\ \hline
    Gowalla & 33,699 & 123,587 & 1,011,694 & 99.98\% & POI \\ \hline
    
    \end{tabular}
    }
    \vspace{-10pt}
\end{table}

%% file: tables/experiment_result.tex
\begin{table*}\centering \footnotesize
\vspace{-10pt}
\renewcommand{\arraystretch}{1}
\caption{Experimental result}
\label{experiment-result-table}
\begin{threeparttable}
\vspace{-10pt}
\resizebox{\linewidth}{!}{
\begin{tabular}{ccccccccccccccc}
\hline
\multicolumn{3}{c}{} & \multicolumn{4}{c}{Epinion} & \multicolumn{4}{c}{MovieLens} & \multicolumn{4}{c}{Gowalla}  \\\cmidrule(lr){4-7}\cmidrule(lr){8-11}\cmidrule(lr){12-15}

\multicolumn{3}{c}{} & Overall & Adv. & Disadv. & \UOFMatric  & Overall & Adv. & Disadv. & \UOFMatric  & Overall & Adv. & Disadv. & \UOFMatric  \\ \hline

\multirow{8}{*}{PMF} & \multirow{4}{*}{NDCG} & Original & 0.3589 & 0.3793 & 0.3578 & 0.0215 & 0.4370 & 0.5097* & 0.4332 & 0.0765 & 0.3756 & 0.4826* & 0.3700 & 0.1126 \\

& & S-DRO & 0.3529 & 0.3652 & 0.3523 & 0.0129 & 0.4379 & 0.4838 & 0.4355 & 0.0483 & 0.3753 & 0.4711 & 0.3703 & 0.1008 \\

& & UFR & \textit{0.3678} & \textit{0.3811} & \textit{0.3671} & \textit{0.0140} & \textit{0.4360} & \textit{0.4901} & \textit{0.4332} & \textit{0.0569} & \textit{0.3747} & \textit{0.4658} & \textit{0.3700} & \textit{0.0958} \\

& & \M & \textbf{0.3804}* & \textbf{0.3861}* & \textbf{0.3801}* & \textbf{0.0060}* & \textbf{0.4638}* & \textbf{0.4809} & \textbf{0.4629}* & \textbf{0.0180}* & \textbf{0.3946}* & \textbf{0.4816} & \textbf{0.3901}* & \textbf{0.0915}* \\ \cline{2-15}

& \multirow{4}{*}{F1} & Original & 0.1050 & 0.1202 & 0.1042 & 0.0160 & 0.1293 & 0.1432* & 0.1286 & 0.0146 & 0.1105 & 0.1240* & 0.1098 & 0.0142
\\

& & S-DRO & 0.1039 & 0.1141 & 0.1034 & 0.0107 & 0.1296 & 0.1388 & 0.1291 & 0.0097 & 0.1104 & 0.1201 & 0.1099 & 0.0102 \\

& & UFR & \textit{0.1055} & \textit{0.1131} & \textit{0.1051} & \textit{0.0080} & \textit{0.1256} & \textit{0.1293} & \textit{0.1254} & \textit{0.0039} & \textit{0.1059} & \textit{0.1104} & \textit{0.1056} & \textit{0.0048} \\

& & \M & \textbf{0.1156*} & \textbf{0.1213}* & \textbf{0.1153}* & \textbf{0.0060}* & \textbf{0.1380}* & \textbf{0.1418} & \textbf{0.1378}* & \textbf{0.0040}* & \textbf{0.1147}* & \textbf{0.1165} & \textbf{0.1146}* & \textbf{0.0019}*
 \\ \hline

\multirow{8}{*}{VAECF} & \multirow{4}{*}{NDCG} & Original & 0.3637 & 0.4000 & 0.3618 & 0.0382 & 0.4404 & 0.5185* & 0.4363 & 0.0822 & 0.4019 & 0.5246 & 0.3768 & 0.1478 \\

& & S-DRO & 0.3623 & 0.3831 & 0.3612 & 0.0219 & 0.4405 & 0.5055 & 0.4371 & 0.0684 & 0.3967 & 0.5029 & 0.3911 & 0.1118\\

& & UFR & \textit{0.3562} & \textit{0.3786} & \textit{0.3550} & \textit{0.0236} & \textit{0.4631} & \textit{0.5124} & \textit{0.4605} & \textit{0.0519} & \textit{0.4005} & \textit{0.4891} & \textit{0.3959} & \textit{0.0932} \\

& & \M & \textbf{0.3927}* & \textbf{0.4012}* & \textbf{0.3923}* & \textbf{0.0089}* & \textbf{0.4783}* & \textbf{0.4858} & \textbf{0.4779}* & \textbf{0.0079}* & \textbf{0.4845}* & \textbf{0.5522}* & \textbf{0.4856}* & \textbf{0.0666}* \\ \cline{2-15}

& \multirow{4}{*}{F1} & Original & 0.1077 & 0.1219* & 0.1070 & 0.0149 & 0.1319 & 0.1476 & 0.1311 & 0.0165 & 0.1157 & 0.1364* & 0.1148 & 0.0216 \\

& & S-DRO & 0.1068 & 0.1200 & 0.1061 & 0.0139 & 0.1316 & 0.1451 & 0.1309 & 0.0142 & 0.1159 & 0.1301 & 0.1152 & 0.0149\\

& & UFR & \textit{0.1038} & \textit{0.1085} & \textit{0.1036} & \textit{0.0049} & \textit{0.1367} & \textit{0.1386} & \textit{0.1366} & \textit{0.0020} & \textit{0.1109} & \textit{0.1136} & \textit{0.1107} & \textit{0.0029}
 \\

& & \M & \textbf{0.1154}* & \textbf{0.1166} & \textbf{0.1153}* & \textbf{0.0013}* & \textbf{0.1476}* & \textbf{0.1505}* & \textbf{0.1474}* & \textbf{0.0031}* & \textbf{0.1315}* & \textbf{0.1353} & \textbf{0.1313}* & \textbf{0.0040}* \\ \hline

\multirow{8}{*}{NeuMF} & \multirow{4}{*}{NDCG} & Original & 0.3606 & 0.3800 & 0.3596 & 0.0204 & 0.4459 & 0.5352 & 0.4412 & 0.0940 & 0.3938 & 0.4717* & 0.3897 & 0.0820 \\ 

& & S-DRO & 0.3598 & 0.3681 & 0.3594 & 0.0087 & 0.4519 & 0.5155 & 0.4486 & 0.0669 & 0.3932 & 0.4529 & 0.3901 & 0.0628\\

& & UFR & \textit{0.3607} & \textit{0.3729} & \textit{0.3600} & \textit{0.0129} & \textit{0.4445} & \textit{0.5068} & \textit{0.4412} & \textit{0.0656} & \textit{0.3935} & \textit{0.4610} & \textit{0.3899} & \textit{0.0711}
\\

& & \M & \textbf{0.4017}* & \textbf{0.4093}* & \textbf{0.4013}* & \textbf{0.0080}* & \textbf{0.5043}* & \textbf{0.5385}* & \textbf{0.5025}* & \textbf{0.0369}* & \textbf{0.3984}* & \textbf{0.4288} & \textbf{0.3968}* & \textbf{0.0320}*
 \\ \cline{2-15}

& \multirow{4}{*}{F1} & Original & 0.1080 & 0.1167 & 0.1075 & 0.0092 & 0.1336 & 0.1432* & 0.1331 & 0.0101 & 0.1153 & 0.1263 & 0.1148 & 0.0115\\

& & S-DRO & 0.1075 & 0.1188 & 0.1069 & 0.0119 & 0.1339 & 0.1401 & 0.1336 & 0.0065 & 0.1149 & 0.1251 & 0.1144 & 0.0107\\

& & UFR & \textit{0.1038} & \textit{0.1085} & \textit{0.1036} & \textit{0.0049} & \textit{0.1302} & \textit{0.1331} & \textit{0.1301} & \textit{0.0030} & \textit{0.1158} & \textit{0.1206} & \textit{0.1156} & \textit{0.0050}
 \\

& & \M & \textbf{0.1194}* & \textbf{0.1213}* & \textbf{0.1193}* & \textbf{0.0020}* & \textbf{0.1416}* & \textbf{0.1432}* & \textbf{0.1417}* & \textbf{0.0015}* & \textbf{0.1269}* & \textbf{0.1307}* & \textbf{0.1267}* & \textbf{0.0040}* \\ \hline

\multirow{8}{*}{NGCF} & \multirow{4}{*}{NDCG} & Original & 0.4200 & 0.4638* & 0.4177 & 0.0461 & 0.5157 & 0.5504* & 0.5139 & 0.0365 & 0.4134 & 0.5179* & 0.4079 & 0.1100 \\

& & S-DRO & 0.4200 & 0.4552 & 0.4181 & 0.0371 & 0.5147 & 0.5479 & 0.5130 & 0.0349 & 0.4108 & 0.5022 & 0.4060 & 0.0962\\

& & UFR & \textit{0.4196} & \textit{0.4462} & \textit{0.4182} & \textit{0.0280} & \textit{0.5221} & \textit{0.5374} & \textit{0.5213} & \textit{0.0161} & \textit{0.4090} & \textit{0.4862} & \textit{0.4049} & \textit{0.0813} \\

& & \M & \textbf{0.4339}* & \textbf{0.4418} & \textbf{0.4335}* & \textbf{0.0083}* & \textbf{0.5370}* & \textbf{0.5422} & \textbf{0.5368}* & \textbf{0.0054}* & \textbf{0.4597}* & \textbf{0.5086} & \textbf{0.4533}* & \textbf{0.0553}* \\ \cline{2-15}

& \multirow{4}{*}{F1} & Original & 0.1203 & 0.1285* & 0.1199 & 0.0086 & 0.1433 & 0.1461 & 0.1432 & 0.0029 & 0.1171 & 0.1333* & 0.1171 & 0.0162 \\

& & S-DRO & 0.1204 & 0.1263 & 0.1201 & 0.0062 & 0.1431 & 0.1396 & 0.1433 & 0.0037 & 0.1166 & 0.1308 & 0.1159 & 0.0130\\

& & UFR & \textit{0.1160} & \textit{0.1208} & \textit{0.1158} & \textit{0.0050} & \textit{0.1420} & \textit{0.1447} & \textit{0.1419} & \textit{0.0028} & \textit{0.1125} & \textit{0.1169} & \textit{0.1122} & \textit{0.0047} \\

& & \M & \textbf{0.1254}* & \textbf{0.1259} & \textbf{0.1254}* & \textbf{0.0005}* & \textbf{0.1469}* & \textbf{0.1472}* & \textbf{0.1469}* & \textbf{0.0003}* & \textbf{0.1284}* & \textbf{0.1322} & \textbf{0.1282}* & \textbf{0.0040}* \\ \hline

\end{tabular}
}
\begin{tablenotes}
        \footnotesize
        \item[*] Note that the italic text indicates the result of UFR. The bold text indicates the result of our proposed \M~framework. The best results are marked with *.
      \end{tablenotes}
    \end{threeparttable}
    \vspace{-10pt}
\end{table*}

%% file: chapters/experiments_and_analysis/dataset_and_experimental_settings.tex
\subsection{Datasets and Experimental Settings}
\nosection{Dataset Description}
We choose three real-world datasets \textbf{Epinion} \cite{massa2007trust}, \textbf{MovieLens} \cite{harper2015movielens}, 
and \textbf{Gowalla} \cite{liu2017experimental} from different domains to fully evaluate the performance of the \M~framework.
The above three datasets have been popularly used \cite{rahmani2022experiments} to prove the performance of models that solve the user-oriented fairness issue.
We list the statistics of the datasets in Table \ref{dataset-statistic}.
Overall, we select these three datasets for three specific reasons in order to demonstrate the scalability, efficiency, and effectiveness of \M. 
\textit{Firstly}, the datasets originate from three distinct domains, including Opinion, Movie, and Point of Interest (POI). 
\textit{Secondly}, the sparsity levels of these datasets vary, which is an important factor that directly impacts the issue of user-oriented fairness, since it determines the activity of users. 
\textit{Lastly}, the three datasets differ in terms of size.
\nosection{Baseline and Backbone Models}
We compare \M~with the state-of-the-art methods UFR \cite{li2021user} and S-DRO~\cite{10.1145/3485447.3512255}.
(1) \textbf{UFR} introduces a post-processing re-rank method to directly modify the recommendation results of a given backbone recommendation model. 
(2) \textbf{S-DRO} provides an in-processing solution that simply minimizes the value of the loss function of disadvantaged users during model training.

%
To fully evaluate the performance of \M, UFR, and S-DRO, we choose four different backbone recommendation models, including a traditional matrix factorization method (PMF), two deep-learning-based methods (VAECF, NeuMF), and a graph neural network-based method (NGCF).
(1) \textbf{PMF} \cite{mnih2007probabilistic}: Probabilistic Matrix Factorization algorithm adds the Gaussian prior into the user and item latent factors distribution to enhance the recommendation performance.
(2) \textbf{VAECF} \cite{liang2018variational}: Variational Autoencoders for Collaborative Filtering proposes a generative model with multinomial likelihood and uses bayesian inference for parameter estimation.
(3) \textbf{NeuMF} \cite{he2017neural}: Neural Collaborative Filtering introduces a deep neural network with non-linear activation functions to train a user and item matching function.
(4) \textbf{NGCF} \cite{wang2019neural}: Neural Graph Collaborative Filtering proposes a graph neural network-based framework to explore the high-order correlations among users and items.
%

%
\nosection{Evaluation Protocols and Parameter Settings}
We utilize the widely-used Leave-One-Out (LOO) \cite{chen2020towards, he2017neural, hu2018conet} strategy to split the train set, tune set, and test set.
Following \cite{li2021user, rahmani2022experiments}, we adopt Normalized Discounted Cumulative Gain (NDCG) and F1-score to evaluate the performance of each model.
A higher value means a better recommendation performance for these two metrics and the predicted cut-off is set as $topK = 10$ \cite{li2021user, rahmani2022experiments}.
Besides, we utilize \UOFMatric~to evaluate the recommendation performance gap between advantaged users and disadvantaged users \cite{li2021user, rahmani2022experiments}.
A smaller \UOFMatric~means a fairer recommendation performance.
We run each model 5 times in each dataset and save the average performance.
Due to space limitations, we introduce the detailed evaluation protocols and parameter settings in Appendix~\ref{appendix-evaluation-protocols} and~\ref{appendix-parameter-settings}.
%

%% file: chapters/experiments_and_analysis/overall_comparison.tex
\vspace{-5pt}
\subsection{Overall Comparison (Q1, Q2)}
We conduct extensive experiments on three datasets based on four backbone models.
The results are reported in Table \ref{experiment-result-table}.
%

%
\nosection{Answer to Q1}
In all these three datasets, models combined with \M~framework significantly outperform the original models.
Overall, by combining the backbone model with our \M~framework, we can not only solve the user-oriented fairness issue but also improve the overall recommendation performance.

\textit{Firstly, \M~can solve the user-oriented fairness issue in the recommendation system.}
%
%
The \UOFMatric~of NDCG and F1 both significantly drop by combining backbone models with \M~framework, proving \M~can narrow the recommendation performance gap between advantaged and disadvantaged users.
We attribute the improvement mainly to the embedding update strategy of \M.
To give disadvantaged users a better learning result in the training process, we extract similar advantaged users for each disadvantaged user by constrained dominant sets approach.
Then, in the training process, we let each disadvantaged user's embedding move closer to the embeddings of advantaged users in the constrained cluster it corresponds to.
Thus disadvantaged users can learn from advantaged users, and the recommendation gap between them is narrowed, leading to a drop in \UOFMatric.

\textit{Secondly, \M~can improve the overall recommendation performance of the backbone model.}
Disadvantaged users are typically the majority group in a recommender system. 
Enhancing the recommendation results for these users can greatly improve overall performance. 
Although the recommendation results for advantaged users can be negatively impacted in some cases, the overall performance is still improved since the percentage of these users is relatively small. 
For instance, when applying the NGCF model combined with \M~to Epinion, the NDCG of advantaged users drops from 0.4638 to 0.4418, while the overall NDCG improves from 0.42 to 0.4339.


\begin{figure}[t]
 \centering
  \includegraphics[width=\linewidth]{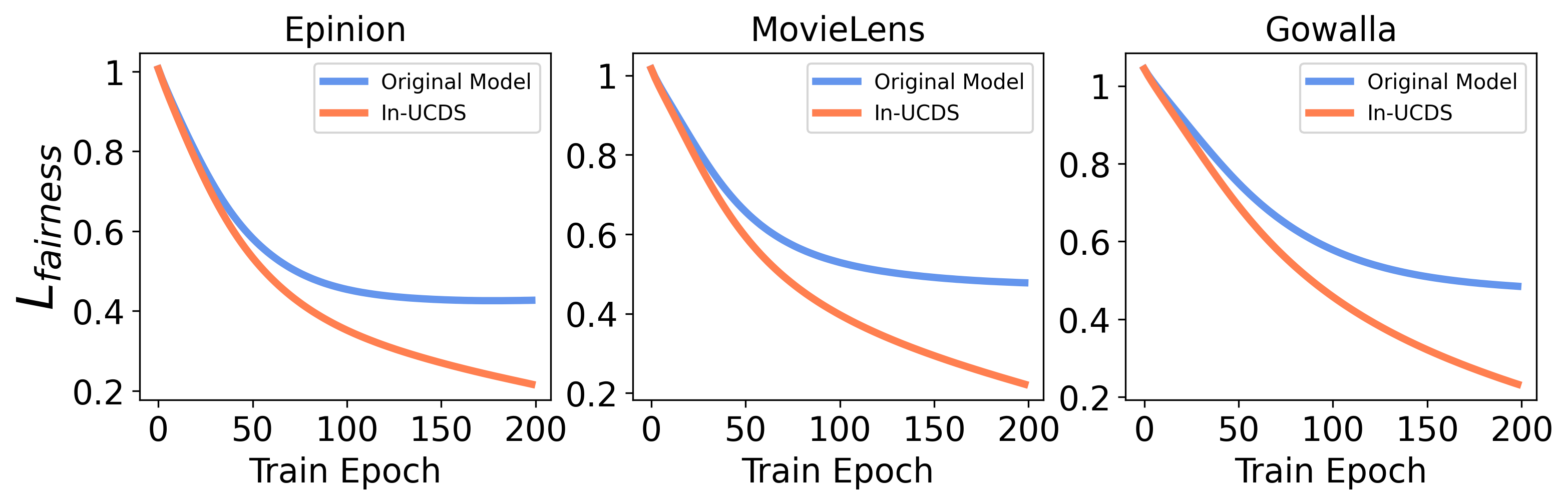}
    \vspace{-20pt}
      \caption{The above results show the change in $L_{fairness}$ as the number of training epochs increases. The result proves that \M~can narrow the learning gap between the advantaged user group and the disadvantaged user group.}
      \vspace{-17pt}
    \label{l-fairness-figure}
\end{figure}

\nosection{Answer to Q2}
\M~has a better performance in all three datasets compared with UFR and S-DRO. 
The superiority of \M~reflects in two aspects.
\textit{Firstly, a better solution to the user-oriented fairness issue.}
As shown in Table \ref{experiment-result-table}, \M~yields consistent best drop of \UOFMatric~on all datasets, proving \M~can better mitigate the recommendation bias between advantaged and disadvantaged users.
%
%
Compared with the post-processing method UFR, we not only consider directly offering a better recommendation performance to disadvantaged users but also make an effort to narrow the bias of the quality of learning results between advantaged and disadvantaged users.
So we can address the essential cause of the user-oriented fairness issue: the learning process of advantaged and disadvantaged users is unfair.
Thus, the gap in the recommendation result between users of different activity levels can be naturally narrowed since the quality of embeddings for disadvantaged users is improved.
In addition, when compared to S-DRO, which focuses on the experience of the worst-case users, \M~aims to enhance the recommendation performance of the entire disadvantaged user group. 
This emphasis on group-level improvement leads to a more substantial enhancement in fairness.

\textit{Secondly, a more significant improvement in overall recommendation performance.}
UFR can only make a limited improvement in the overall recommendation performance; sometimes, it even has a negative effect.
The reason is that UFR directly re-ranks the recommendation result to achieve user-oriented fairness.
However, without enhancing the backbone model's training process, the re-rank method's result is unstable.
%
%
The objective of S-DRO is to minimize the loss for the worst-performing group.
However, due to the inherent limitations of the training data for disadvantaged users, it is challenging to achieve significant improvement.
In contrast to the aforementioned methods, \M~has the advantage of directly enhancing the training process of disadvantaged users by leveraging the learning from advantaged users. 
This approach is more intuitive and natural, ensuring a positive impact on the model's performance.
Besides, the performance of advantaged users is improved occasionally since they can also learn from disadvantaged users who have stronger embeddings.
For example, the NDCG of advantaged users in Epinion for the PMF model improves from 0.3793 to 0.3861, reflecting improved learning results.

\nosection{Model Efficiency}
Moreover, we compare the training time of the original backbone models, UFR, S-DRO, and \M~in Appendix~\ref{appendix-model-efficiency} to prove that \M~can keep the time cost at the same level as the original model.

%% file: chapters/experiments_and_analysis/the_change_in_l_fairness.tex
\subsection{The Change in $L_{fairness}$ (Q3)}


%
In this part, we choose NeuMF as the example backbone model since it is widely used to evaluate the user-oriented fairness model \cite{li2021user, rahmani2022experiments}.
To prove that \M~can narrow the gap in the quality of learned embeddings between disadvantaged users and advantaged users, we show the change of $L_{fairness}$ with the number of training epochs increases in Figure~\ref{l-fairness-figure}.
Note that $L_{fairness}$ can directly reflect the difference between embeddings of two user groups as Equation \eqref{D-overall-equ} shows.
%
%
%
The experimental results indicate that the model combined with \M~can progressively narrow the difference between disadvantaged and advantaged users, thus significantly improving the learning process of the former by learning from similar advantaged users.

%% file: chapters/experiments_and_analysis/effect_of_hyperparameter_k.tex
\subsection{Effect of Hyperparameters (Q4)}
%
We show the effect of the most important hyperparameters $k$ (see Equation~\eqref{D-equ-new}) and $\beta$ (see Equation~\eqref{loss-overall}) of \M~in Appendix~\ref{appendix-effect-of-hyperparameters}. 
%
%
Through the experimental results, we set $k$ as 3 and $\beta$ as 0.00001.
%



%% file: chapters/experiments_and_analysis/ablation_study.tex
\subsection{Ablation Study (Q5)}
\input{tables/ablation_study_table}
In this section, we choose NeuMF as the example backbone model to perform ablation experiments to demonstrate the efficiency of UCDS. 
We introduce a simplified version called In-Naive for comparative analysis. 
%
%
Specifically, for each disadvantaged user, we rank the advantaged users based on the number of same interactions with him/her and extract the top-$k$ advantaged users.
The results presented in Table~\ref{ablation-study-table}~demonstrate that In-UCDS consistently outperforms In-Naive, which highlights the importance of incorporating UCDS into our framework.
This is because the number of similar interactions alone may not accurately reflect the true similarity among users.
A great deal of the same interactions come from interactions with common items, which do not reflect the characteristics and uniqueness of the user.
In contrast to In-Naive, by leveraging the graph structure, UCDS can uncover deep relevance and some high-order correlations among users.
The discovered user clusters are more reflective of the uniqueness of each disadvantaged user, which helps prevent disadvantaged users from overfitting to common items.

%% file: tables/ablation_study_table.tex
\begin{table}\centering \footnotesize
\renewcommand{\arraystretch}{1.1}
  \vspace{-6pt}
  \caption{Ablation study}
  \vspace{-8pt}
  \label{ablation-study-table}
  \resizebox{\linewidth}{!}{
    \begin{tabular}{cccccccc}
    \hline
& & \multicolumn{2}{c}{Epinion} & \multicolumn{2}{c}{MovieLens} & \multicolumn{2}{c}{Gowalla}  \\\cmidrule(lr){3-4}\cmidrule(lr){5-6}\cmidrule(lr){7-8}

& & Overall & \UOFMatric  & Overall & \UOFMatric  & Overall & \UOFMatric  \\
\hline

\multirow{2}{*}{NDCG} & In-Naive & 0.3783 & 0.0159 & 0.4799 & 0.0558 & 0.3932 & 0.0655\\

& \M & \textbf{0.4017} & \textbf{0.0080} & \textbf{0.5043} & \textbf{0.0369} & \textbf{0.3984} & \textbf{0.0320} \\ \hline

\multirow{2}{*}{F1} & In-Naive & 0.1100 & 0.0051 & 0.1339 & 0.0047 & 0.1198 & 0.0088\\

& \M & \textbf{0.1194} & \textbf{0.0020} & \textbf{0.1416} & \textbf{0.0015} & \textbf{0.1269} & \textbf{0.0040} \\

\hline 
    
    \end{tabular}
    }
    \vspace{-11pt}
\end{table}

%% file: chapters/conclustion.tex
\section{Conclusion}
In this paper, we focus on rarely studied User-Oriented Fairness (UOF) in recommender systems, which aims to narrow the recommendation performance gap between advantaged and disadvantaged users.
We propose an \textbf{In}-processing \textbf{U}ser \textbf{C}onstrained \textbf{D}ominant \textbf{S}ets (\M) approach, which can be divided into two stages, i.e., the \textit{\Mfirst~modeling stage} and the \textit{in-processing training stage}.
Through the above two stages, we let disadvantaged users learn from their similar advantaged users to narrow the recommendation performance gap between these two groups of users.
We conduct extensive experiments on three real-world datasets based on four different backbone models.
The experimental results prove that \M~can deal with the UOF issue better than the state-of-the-art methods and provide a better overall recommendation performance.

%% file: chapters/appendix.tex
\appendix

\begin{figure*}[htbp]
    \centering
    \subfigure[User Graph Example]{
		\label{toy1-figure}		\includegraphics[width=0.155\linewidth]{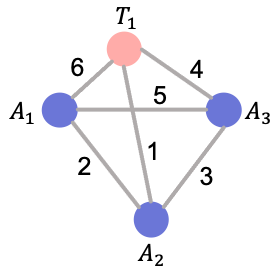}}
     \subfigure[$w_{\{T_1, A_1, A_2, A_3\}}(A_2) < 0$]{
		\label{toy2-figure}
		\includegraphics[width=0.155\linewidth]{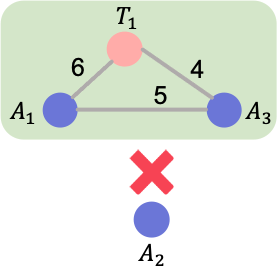}}
     \subfigure[$w_{\{T_1, A_1, A_3\}}(A_3) > 0$]{
		\label{toy3-figure}
		\includegraphics[width=0.155\linewidth]{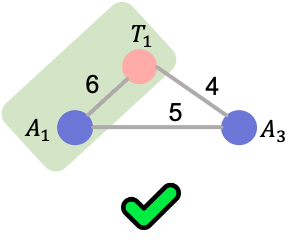}}
   \subfigure[$w_{\{T_1, A_1, A_3\}}(A_1) > 0$]{
		\label{toy4-figure}
		\includegraphics[width=0.155\linewidth]{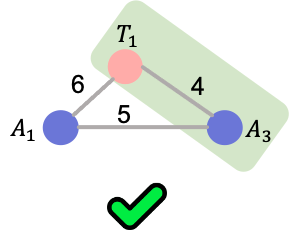}}
   \subfigure[$w_{\{T_1, A_1, A_3\}}(T_1) > 0$]{
		\label{toy5-figure}
		\includegraphics[width=0.155\linewidth]{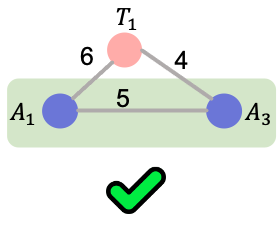}}
   \subfigure[\textbf{Coherent Cluster $\mathcal{S}$}]{
		\label{toy6-figure}
		\includegraphics[width=0.155\linewidth]{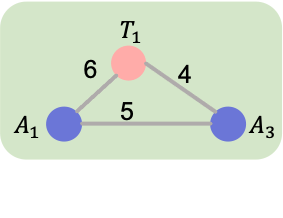}}
  
    \caption{A toy example to show how to construct a user-constrained cluster. The example graph is extracted from Figure 3 in the main paper and the disadvantaged user $T_1$ is the target user. The weight of each edge is set as the number of the same items that linked two users have both interacted with in the user-item interaction graph in Figure 3 in the main paper.
    }
    \label{toy-example-figure}
\end{figure*}
\begin{figure*}[htbp]
  \centering
  \includegraphics[width=\linewidth]{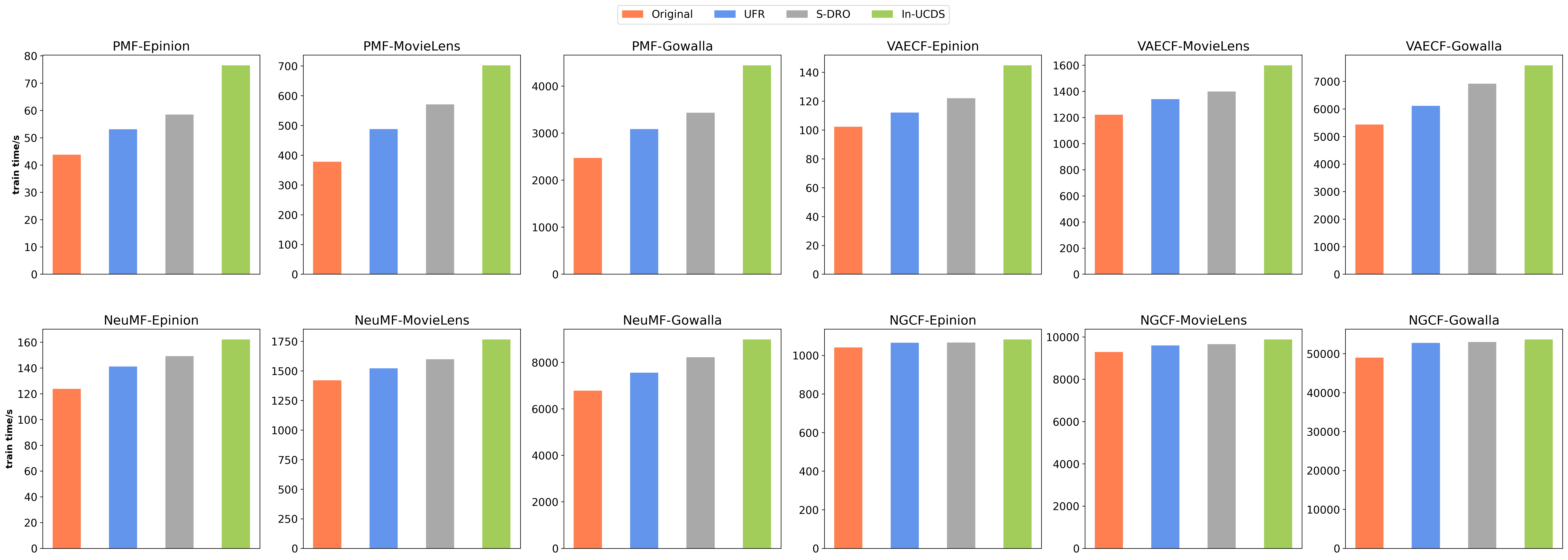}
      \caption{The training time of four original backbone models, models combined with UFR,  models combined with S-DRO, and models combined with \M~in three datasets.}
      \label{model-efficiency-figure}
\end{figure*}

\begin{figure}[t]
  \centering
  \includegraphics[width=\linewidth]{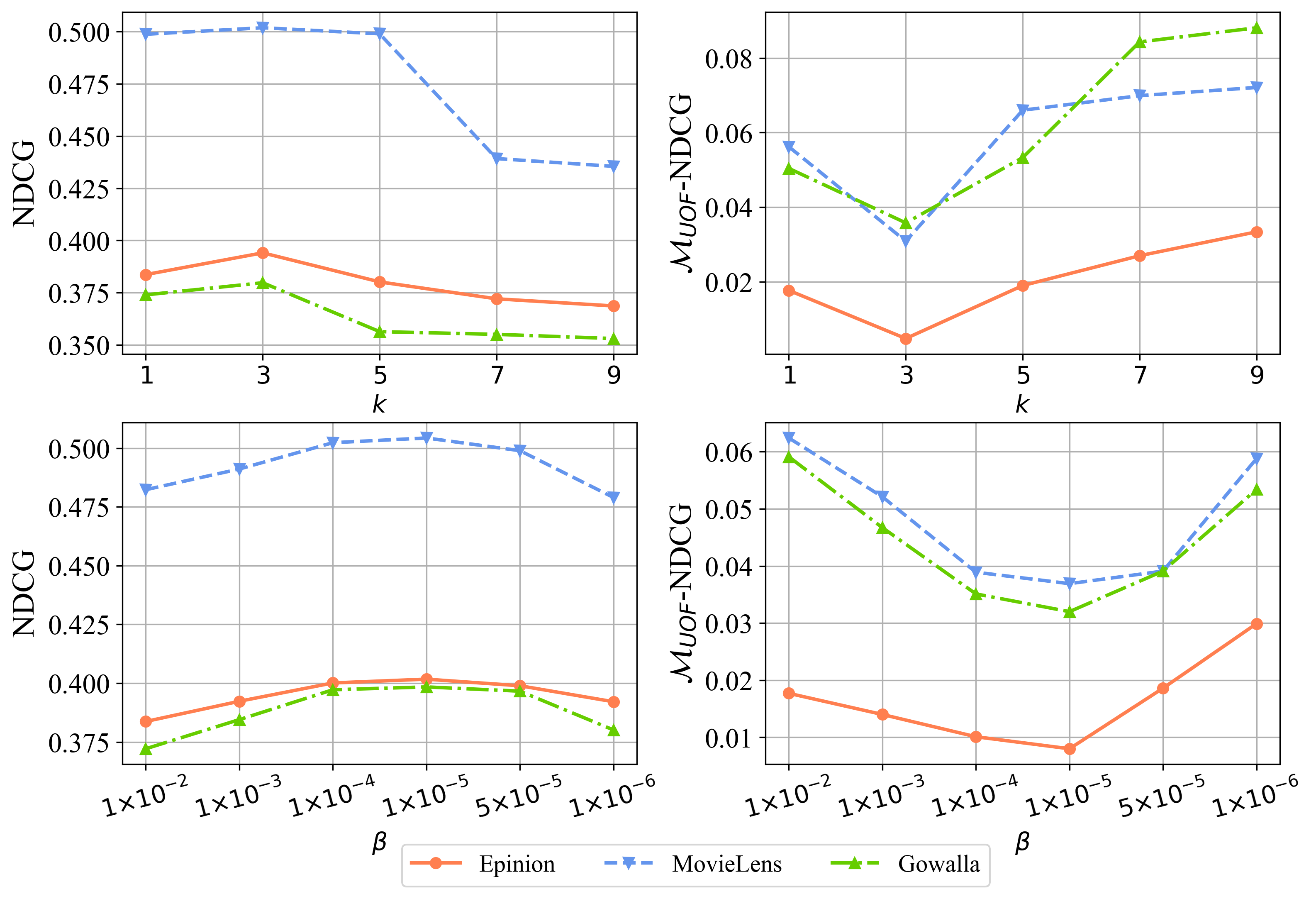}
      \caption{The effect of hyperparameters $k$ and $\beta$. We show the fluctuations of NDCG and \UOFMatric-NDCG with the value of $k$ and $\beta$.
      }
      \label{effect-of-k-figure}
\end{figure}

\section{Example of constructing a user-constrained dominant set \label{appendix-example}}

We give a toy example in Figure \ref{toy-example-figure}~to describe how we extract a dominant set in detail.
%
%
Given a user-user graph in Figure \ref{toy1-figure},
adding $A_2$ to the user sets will reduce the coherence, as Figure \ref{toy2-figure} shows. 
Thus, $A_2$ is removed from $\mathcal{S}$.
In Figure \ref{toy3-figure}, \ref{toy4-figure}, \ref{toy5-figure}, since all users of $\{T_1, A_1, A_3\}$ have a positive effect on the coherence of the cluster, they are kept in the final cluster in Figure \ref{toy6-figure}.

\section{More Experimental Details \label{appendix-more-experimental-details}}

\subsection{Evaluation Protocols \label{appendix-evaluation-protocols}}
According to the experiment results of \cite{li2021user, rahmani2022experiments}, the number of interactions gives the most intuitive indication of which group (i.e., advantaged or disadvantaged users) a user belongs to.
Thus, following \cite{li2021user, rahmani2022experiments}, we rank users in each dataset according to their number of interactions and extract the top 5\% users as advantaged users, leaving other users as disadvantaged users.

How to divide the experimental datasets is important for user-oriented fairness.
Existing work \cite{li2021user, rahmani2022experiments} always splits each dataset as the train set of 80\%, the validation set of 10\%, and the test set of 10\%.
They select 100 negative items for each user in the validation and test process to calculate the rank result combined with their positive samples.
We claim that the gap in the recommendation performance of advantaged users and disadvantaged users is artificially magnified by doing so.
The reason is that by splitting the datasets as described above, the number of positive items of advantaged users is much larger than that of disadvantaged users.
Then, with the same number (100) of negative items, the proportion of the positive samples in the test sets of advantage users is much larger than that of disadvantaged users.
In this situation, advantaged users are more likely to get satisfied recommendation results.
This is contrary to the real-life situation where the number of negative samples in a recommender system is much greater than the number of positive samples.
To overcome such bias, we utilize the widely-used Leave-One-Out (LOO) \cite{chen2020towards, he2017neural, hu2018conet} strategy to do the dataset splitting proces.
In detail, each validation set and test set for a user contains one randomly selected positive item and 99 negative items.

Following \cite{li2021user, rahmani2022experiments}, we adopt Normalized Discounted Cumulative Gain (NDCG) and F1-score to evaluate the performance of each model.
A higher value means a better recommendation performance for these two metrics and the predicted cut-off is set as $topK = 10$ \cite{li2021user, rahmani2022experiments}.
Besides, we utilize \UOFMatric~to evaluate the recommendation performance gap between advantaged users and disadvantaged users \cite{li2021user, rahmani2022experiments}.
A smaller \UOFMatric~means a fairer recommendation performance.
We run each model 5 times in each dataset and save the average performance.

\subsection{Parameter Settings \label{appendix-parameter-settings}}
\textit{For \M}, we set the hyperparameter $k$ as 3 and $\beta$ as 0.00001 according to the experimental result of hyperparameters. (see section 5.4 in the main paper).
We tune the value of $\alpha$ following Equation (9) in the main paper for each dataset.
\textit{For UFR}, we use the code provided by authors and leave the parameters as their default values.
\textit{For S-DRO}, we implement it by ourselves. As recommended in \cite{10.1145/3485447.3512255}, we set the dimension of hidden layers as (128, 64) and temperature $\tau$ as 0.07.
\textit{For backbone models}, we set the size of user and item embeddings to 64 for all of them, and adopt their parameters as suggested in their original paper.
We apply Adam \cite{kingma2014adam} as the optimizer and set the learning rate to 0.0001.
To ensure the convergence of all models, we set the number of training epochs to 200.


\subsection{Model Efficiency \label{appendix-model-efficiency}}

In-processing methods always require additional operations to achieve fairness, leading to potential impacts on the training time of the backbone model.
Therefore, it is necessary to evaluate the efficiency of the \M~method. 
To demonstrate this, we showcase the training time of the original backbone model, the model combined with UFR, the model combined with S-DRO, and the model combined with \M~in Figure \ref{model-efficiency-figure}. 
To ensure consistency, we fix the number of training epochs as 200 and calculate the average training times of 5 runs for each model.

Overall, the \M~method requires the longest training time compared to the other methods, but we can keep the time cost at the same level as the original model. 
\M~needs to extract an advantaged user cluster for each disadvantaged user before training the model, which is more complex than the re-ranking goal of UFR and the simple optimization process of S-DRO, causing it to be the most time-consuming method.
However, the time cost of the \Mfirst~modeling stage is independent of the complexity of the backbone model and is only relevant to the size of the dataset.
Therefore, the time cost of \M~becomes more insignificant as the complexity of the backbone model increases. 
For example, the proportion of extra time-consuming in NGCF is much smaller than that in PMF.

\subsection{Effect of Hyperparameters \label{appendix-effect-of-hyperparameters}}
In this part, we show the effect of the most important hyperparameters $k$ (see Equation~\eqref{D-equ-new}) and $\beta$ (see Equation~\eqref{loss-overall}) of \M. 
We take NeuMF as the example backbone model to show the experimental results in Figure \ref{effect-of-k-figure}.

\nosection{Effect of $k$}
Through the experimental results, we can find that when the value of $k$ is 3, the model achieves a high level of both recommendation performance and fairness.
Such a phenomenon shows that by exploring 3 similar advantaged users for each disadvantaged user, \M~can enhance the learning result of disadvantaged users and maintains their characteristics simultaneously.
%
%
It should be noticed that the performance and fairness of the model both decrease when the value of $k$ is too large.
The reason is that if a constrained cluster contains too many users, the unique characteristics of a cluster and its corresponding disadvantaged user will be neutralized, negatively affecting the training process.
%

\nosection{Effect of $\beta$}
Overall, \M~can achieve the best recommendation performance and fairness when $\beta = 0.00001$.
%
%
It should be noticed that the model's recommendation performance and fairness follow the same trend as the change in $\beta$ since they are both affected by the quality of the learning results of disadvantaged users.